\documentclass[aps,prb,twocolumn,reprint,amsmath,amssymb,superscriptaddress]{revtex4-2}

\usepackage{graphicx}% Include figure files
\usepackage{dcolumn}% Align table columns on decimal point
\usepackage{bm}% bold math
\usepackage{multirow}
\usepackage{subfigure}
\usepackage{booktabs}
\usepackage{hhline}  % Double line tablet
\usepackage{float}
\usepackage{braket}
\usepackage[colorlinks,linkcolor=blue,anchorcolor=blue,citecolor=blue,urlcolor=blue]{hyperref}

\begin{document}

%\preprint{}

\title{First-principles estimation of the superconducting transition temperature of a metallic hydrogen liquid}


\author{Haoran Chen}
\affiliation{International Center for Quantum Materials, Peking University, Beijing 100871, China}

\author{Xiao-Wei Zhang}
\affiliation{International Center for Quantum Materials, Peking University, Beijing 100871, China}

\author{Xin-Zheng Li}
\affiliation{State Key Laboratory for Artificial Microstructure and Mesoscopic Physics, and School of Physics, Peking University, Beijing 100871, China}

\author{Junren Shi}
\email{junrenshi@pku.edu.cn}
\affiliation{International Center for Quantum Materials, Peking University, Beijing 100871, China}
\affiliation{Collaborative Innovation Center of Quantum Matter, Beijing 100871, China}


\date{\today}

\begin{abstract}
	We present a full density-functional-theory-based implementation of the stochastic path-integral approach proposed by Liu et al~.\cite{liu_superconducting_2020} % [Liu et al., Phys. Rev. Research 2, 013340 (2020)]
	for estimating the superconducting transition temperature ($T_c$) of a liquid. The implementation is based on the all-electron projector-augmented-wave (PAW) method. We generalize Liu et al.'s formalism to accommodate the pseudo-description of electron states in the PAW method. A  formula for constructing the overlap operator of the PAW method is proposed to eliminate errors due to the incompleteness of a pseudo-basis set. We apply the implementation to estimate $T_c$'s of metallic hydrogen liquids.  It confirms Liu et al.'s prediction that metallic hydrogen could form a superconducting liquid.
\end{abstract}
%\keywords{}

\maketitle

\section{Introduction}
Ever since the discovery of the superconductivity by K. Onnes in 1911, the search for materials with high superconducting transition temperatures ($T_c$'s) had been a constant pursuit.  The recent discovery of a new family of high-$T_c$ superconductors in hydrides under high pressure shows the great promise of first-principles calculations in guiding the search~\cite{duan_pressure-induced_2014,liu_potential_2017}. Based on the celebrated Bardeen-Copper-Schrieffer (BCS) theory of superconductivity, it was long predicted that metallic hydrogen or hydrogen-rich materials could exhibit high $T_c$'s as light hydrogen atoms provide both a high Debye frequency of phonons and strong electron-phonon coupling (EPC)~\cite{ashcroft_metallic_1968,jaffe_superconductivity_1981,richardson_high_1997,mcmahon_high-temperature_2011}.  By using state-of-art first-principles approaches based on the density-functional theory (DFT), it is now possible to conduct an extensive search in various hydrides, and predict their structures and  $T_c$'s~\cite{duan_pressure-induced_2014,liu_potential_2017,kim_predicted_2011,flores-livas_superconductivity_2016,sun_route_2019}.

In spite of the tremendous success, the first-principles prediction of $T_c$ for compounds like hydrides is tricky.  The widely-adopted algorithm is based on the Eliashberg theory which relates $T_c$ to quantities calculable by DFT such as the band structure of electrons and the matrix elements of EPC~\cite{grimvall_electron-phonon_1981,giustino_electron-phonon_2017}. The theory is built upon the assumption that the vibration amplitudes of atoms in a solid are small. It thus adopts the harmonic approximation for describing atom vibrations and a perturbative approach for determining EPC effects.  In hydrides, however, the assumption likely breaks down. Light hydrogen atoms tend to have large vibration amplitudes, and therefore their vibrations are subject to anharmonicity~\cite{errea_high-pressure_2015}.  Moreover, hydrogen atoms can tunnel between lattice sites and could even form a superionic phase in which they become delocalized and diffuse in a whole lattice~\cite{liu_dynamics_2018}.  In the mother compound of metallic hydrides, i.e., the metallic hydrogen, quantum fluctuations induced by the tunneling is so strong that its melting point is predicted to be suppressed below the room temperature~\cite{chen_quantum_2013,geng_lattice_2015}.  The applicability of the Eliashberg theory in these systems is questionable.

To date, most of the efforts addressing the quantum tunneling and anharmonic effects in superconductors are based on the approach of the stochastic self-consistent harmonic approximation (SSCHA)~\cite{errea_anharmonic_2014} which treats an anharmonic system as an effective harmonic one with self-consistently determined phonon modes.  $T_c$ can then be estimated by applying the standard algorithm for a harmonic system.  SSCHA calculations show that the anharmonicity and quantum fluctuations could yield different predictions of crystal structures and $T_c$'s~\cite{errea_high-pressure_2015,errea_quantum_2020}.  While the approach is effective and efficient, it relies on a cascade of approximations which could become uncontrolled.  In particular, when a system enters a liquid or superionic phase, the underlying assumption that the system can be approximated as a harmonic solid obviously breaks down.

Recently, Liu et al. propose a stochastic path-integral approach of determining $T_c$ for general systems including liquids~\cite{liu_superconducting_2020}. The core of the approach is a set of rigorous relations through which an effective pairing interaction between electrons can be inferred from the fluctuations of electron-ion scattering $\mathcal{T}$-matrices. The fluctuations of the $\mathcal{T}$-matrices can be estimated from a stochastic sampling of ion positions generated by a path integral molecular dynamics (PIMD) simulation~\cite{marx_ab_1996}.  Furthermore, it is shown that the equations for determining $T_c$ in the Eliashberg theory, i.e., the linearized Eliashberg equations, can be reestablished rigorously in the new context. As a result, with the effective pairing interaction, $T_c$ can be determined in the exactly same manner as in the conventional Eliashberg approach.  The approach is based on a rigorous theory and makes no assumption on the nature of atom motion. It should therefore apply equally well for solids or liquids, with or without anharmonicity and quantum fluctuations.

Liu et al. present an implementation of the approach for metallic hydrogen liquids in Ref.~\onlinecite{liu_superconducting_2020}.  They predict that metallic hydrogen would form a superconducting liquid with $T_c$ well above its melting temperature. In the implementation, hydrogen atom positions are sampled by using the first-principles PIMD method.  However, ionic fields, i.e., the interaction potentials between hydrogen ions and electrons, are approximated as a bare Coulomb potential screened by a dielectric function.  Such a linear screening approximation (LSA) should be adequate for metallic hydrogen because hydrogen ions, i.e., protons, are point particles without inner-core electrons.  For more general systems involving other atom species as in the case of hydrides, however, the approximation is obviously inadequate.  The deficiency limits the applicability of the implementation to metallic hydrogen only.

In this paper, we develop a full DFT-based implementation of the stochastic path-integral approach for determining $T_c$.  We base our implementation on the Vienna Ab-initio Simulation Package (VASP) \cite{kresse_efficiency_1996,kresse_efficient_1996}, and determine ionic fields by employing the projector augmented-wave (PAW) method \cite{blochl_projector_1994,kresse_ultrasoft_1999}.  In the method, the states of valance electrons are described by pseudo (PS) wavefunctions, and the all-electron (AE) wavefunctions, i.e., the true wavefunctions of electrons, are related to PS-wavefunctions through a transformation which depends on the positions of ions.  It thus introduces complexities in directly applying Liu et al.'s formalism. We solve the issue by generalizing the formalism and accommodating it to the PS description. The generalized formalism enables our full DFT-based implementation.  As a test, we apply the implementation to metallic hydrogen liquids.  It yields results close to Liu et al.'s results, confirming the prediction that metallic hydrogen could form a superconducting liquid.

The remainder of the paper is organized as follows. In Sec.~\ref{Theory}, we outline the formalism and calculation procedure of the stochastic path-integral approach proposed by Liu et al., and identify main approximations in their implementation. In Sec.~\ref{PAW2AE}, we generalize and implement Liu et al.'s formalism to accommodate the PAW method.  A number of issues relevant to the implementation are also addressed. In particular, a formula for constructing the overlap operator in the PAW method is proposed to eliminate errors due to the incompleteness of PS basis set. In Sec.~\ref{Result}, we apply the implementation to metallic hydrogen liquids, and compare results with Liu et al.'s results.  The convergency of the calculation is also tested. Finally, we summarize our results in Sec.~\ref{Summary}.

%\section{Theory}\label{Theory}
\section{Stochastic path-integral approach of determining $T_c$}\label{Theory}
In this section, we outline the formalism of the stochastic path-integral approach of determining $T_c$ proposed by Liu et al. and discuss approximations involved in implementing the approach.

\subsection{Formalism}\label{G_theory}
In the conventional Eliashberg theory, $T_c$ is determined by solving a set of linearized Eliashberg equations.  Liu et al. show rigorously that the same set of the equations still hold for liquids and general systems, provided that relevant parameters are properly determined. The set of the linearized Eliashberg equations reads
\begin{eqnarray} 
	\rho\Delta_n &=& \sum_{n'}
	\left[
		\lambda(n'-n) - \mu^\star - \frac{\hbar\beta}{\pi}
		|\tilde{\omega}(n)| \delta_{nn'}
		\right]
	\Delta_{n'},\quad \label{Eliash01}\\
	\tilde{\omega}(n) &=& \frac{\pi}{\hbar\beta}
	\left(
	2n + 1 + \lambda(0) + 2\sum_{m=1}^n \lambda(m)
	\right), n \ge 0\quad\label{Eliash02}
\end{eqnarray}
and $|\tilde{\omega}(-n)|=|\tilde{\omega}(n-1)|$, where $n$ and $n^\prime$ are integers indexing the Matsubara frequencies of Fermions $\omega_n=(2n+1)\pi/\hbar\beta$, $\beta \equiv 1/k_B T$ and $T$ is the temperature.  $\{\lambda(m),\, m \in Z\}$ is a set of interaction parameters characterizing the effective pairing interaction, and $\mu^*$ is the empirical Morel-Anderson pseudopotential introduced to take account of the exchange-correlation effect of electrons \cite{morel_calculation_1962}. Solving Eq.~(\ref{Eliash01}) yields a set of eigenvalues $\rho$ and eigenvectors $\Delta_n$. For normal states, all eigenvalues are negative. The emergence of a non-negative eigenvalue indicates that the system enters into the superconducting state. One can thus estimate $T_c$ by solving the equations for a number of temperatures to see  when the maximal eigenvalue changes sign. We note that $\Delta_n$ here is not related to the physical gap.

Two comments are in order.  First, in calculations for solids using the standard approach (see, e.g., Ref.~\onlinecite{mcmahon_high-temperature_2011}), it is common to use the empirical McMillan formula and estimate $T_c$ from a small number of parameters evaluated at the zero temperature including the mass enhancement factor $\lambda \equiv \lambda(0)$, an average phonon frequency and $\mu^*$.  It simplifies the calculations. Unfortunately, such an approach does not work well for liquids because (a) unlike harmonic solids, $\lambda$ and the average phonon frequency of a liquid are temperature-dependent; (b) the empirical formulas are tuned for solids, and do not fit well with liquids which have qualitatively different density spectral functions due to the presence of low-frequency diffusive modes~\cite{knorr_superconductivity_1970,bergmann_eliashberg_1971}.  Second, the choice of parameter $\mu^*$ is empirical.  For metallic hydrogen systems,  we use $\mu^* \approx 0.089$ as suggested in Ref.~\onlinecite{mcmahon_high-temperature_2011}.  The empirical parameter introduces an uncertainty in predicting $T_c$.  Ideally, one should base calculations on the DFT for superconductors~\cite{gross_density-functional_1991,luders_ab_2005,marques_ab_2005}, which provides a rigorous framework for treating exchange-correlation effects in superconductors.  This is not yet implemented in our current calculation.

To apply the Eliashberg equations, the interaction parameters $\lambda(m)$ are the central quantities to be determined. They are related to the effective pairing interaction $W$ by
\begin{eqnarray} \label{Eliash03}
	\lambda(n-n^\prime)=-\sum_{\bm{k}^\prime}
	W(\omega_{n}-\omega_{n^\prime}, \bm k-\bm k^\prime)\delta(\tilde{\epsilon}_{\bm{k}^\prime}-\epsilon_F),
\end{eqnarray}
where $W(\omega_{n}-\omega_{n^\prime}, \bm k-\bm k^\prime)$ denotes the matrix element of the effective pairing interaction, which is a function of the quasi-wave-vector transfer $\bm k - \bm k^\prime$ and the frequency transfer $\omega_n - \omega_{n^\prime}$, $\tilde \epsilon_{\bm k^\prime}$ is the renormalized quasi-electron dispersion, and $\epsilon_F$ is the Fermi energy.  To apply the formula, it is necessary to determine the effective pairing interaction as well as the renormalized quasi-electron dispersion.

To determine these quantities in a liquid, Liu et al. propose a stochastic path integral approach.  It is based on the classical isomorphism \cite{chandler_exploiting_1981}, i.e.,  mapping quantum ions to classical ring polymers by applying the imaginary-time path integral formalism~\cite{negele_quantum_1988}.  The spatial configuration of the classical ensemble of ring polymers is specified by a set of $\tau$-dependent coordinates $\{\bm R_a(\tau)\}$, where $\tau \in [0,\hbar\beta)$ is the imaginary time or interpreted as a variable to parameterize the rings.  In a PIMD simulation, the isomorphic ensemble is simulated by using the classical molecular dynamics.  It provides a statistical sampling of the spatial configurations of the ring polymers.

Quantities related to electrons can be evaluated in the isomorphic ensemble.  Electrons in the ensemble experience a $\tau$-dependent Kohn Sham potential $V_\mathrm{KS}(\tau)$ which is self-consistently determined by the ion configuration $\{\bm R_a(\tau)\}$ at the specific imaginary-time $\tau$.  As a result, the single-particle Green's function of electrons with respect to a given configuration of ring polymers, in the context of the Kohn-Sham theory, can be determined by
\begin{equation}
	\left[- \frac{\partial}{\partial\tau}-\frac{\hat{H}(\tau)-\epsilon_F\hat{\mathbb{I}}}{\hbar}\right] \hat{\mathcal{G}}(\tau,\tau^\prime) = \delta(\tau-\tau^\prime )\hat{\mathbb{I}},\label{GreenG}
\end{equation}
where we write the equation in an operator form, $\hat{H}(\tau)\equiv \hat{K}+\hat{V}_{\mathrm{KS}}(\tau)$
with $\hat{K}$ being the operators of the kinetic energy.  The Green's function is not a physical Green's function for electrons in a liquid.  To get the physical one, we need to further conduct an ensemble average:
%with $\hat{K}$ and $\hat V_{\mathrm{KS}}$ being the operators of the kinetic energy and the self-consistent Kohn-Sham potential, respectively.  The Green's function is not a physical Green's function for electrons in a liquid.  To get the physical one, we need to further conduct an ensemble average:
\begin{eqnarray}
	\mathcal{G}_\mathrm{phy.}\equiv \bar{\mathcal{G}}(\tau-\tau') = \langle\mathcal{G}(\tau,\tau')\rangle_C,
\end{eqnarray}
where $\langle\dots\rangle_C$ denotes the average over the ring-polymer configurations.  The averaged (physical) Green's function $\bar{\mathcal{G}}(\tau-\tau^\prime)$ defines the dispersion and lifetime of a quasi-electron renormalized by the fluctuating ionic field.  The quasi-electron dispersion $\tilde \epsilon_{\bm k^\prime}$ required by Eq.~(\ref{Eliash03}) can then be inferred from it. 

For a given spatial configuration of ring polymers, we can determine the electron-ion scattering $\mathcal{T}$-matrix. The $\mathcal{T}$-matrix is related to the Green's function according to the Lippmann-Schwinger equation
\begin{eqnarray} \label{LSE}
	\hat{\mathcal{T}} = 
	\hat{\mathcal{V}} + 
	\frac{1}{\hbar} \hat{\mathcal{V}}
	\hat{\bar{\mathcal{G}}}
	\hat{\mathcal{T}},
\end{eqnarray}
where $\hat{\mathcal{V}}=\hat{V}_\mathrm{KS}(\tau) - \hat{\bar\Sigma}$ is the effective scattering potential for electrons, and $\hat{\bar\Sigma}$ is the self-energy inferred from $\hat{\bar{\mathcal{G}}}$.

%\begin{equation}
%	\hat{\mathcal{T}}=\hbar\hat{\bar{\mathcal{G}}}^{-1}\hat{\mathcal{G}}\hat{\bar{\mathcal{G}}}^{-1}-\hbar\hat{\bar{\mathcal{G}}}^{-1}.\label{Tmatrix}
%\end{equation}

%For a given spatial configuration of ring polymers, we can also determine the  $\mathcal{T}$-matrix of an electron scattered by the ionic field.  It is defined by the operator identity
%\begin{equation}
%	\hat{\mathcal{T}}=\hbar\hat{\bar{\mathcal{G}}}^{-1}\hat{\mathcal{G}}\hat{\bar{\mathcal{G}}}^{-1}-\hbar\hat{\bar{\mathcal{G}}}^{-1},\label{Tmatrix}
%\end{equation}
%and satisfies the Lippmann-Schwinger equation
%\begin{eqnarray} \label{LSE}
%	\hat{\mathcal{T}} = 
%	\hat{\mathcal{V}} + 
%	\frac{1}{\hbar} \hat{\mathcal{V}}
%	\hat{\bar{\mathcal{G}}}
%	\hat{\mathcal{T}},
%\end{eqnarray}
%where $\hat{\mathcal{V}}=\hat{V}_{\mathrm{KS}}[\rho] + \hat{V}_\mathrm{ion}(\tau) - \hat{\bar\Sigma}$ is the effective scattering potential for electrons, and $\hat{\bar\Sigma}$ is the self-energy inferred from $\bar{\mathcal{G}}$.

Liu et al. establish a rigorous relation between the effective pairing interaction and the $\mathcal{T}$-matrices. One first determines the pair scattering amplitude $\Gamma$, i.e., the fluctuation of the $\mathcal{T}$-matrices:
\begin{eqnarray} \label{scatter}
	\Gamma_{11'} = -\beta 
	\langle | \mathcal{T}_{11'}| ^2 \rangle _C,
\end{eqnarray}
where $\mathcal{T}_{11'}$ denotes the matrix element of the $\mathcal{T}$-matrix, and we adopt a short-hand notation by using numbers to represent the states of electrons: $1\equiv (\bm{k},\omega_n)$, $\bar 1 \equiv (-\bm{k},-\omega_n)$, and similarly for $1^\prime$ and $\bar 1^\prime$.  Then, the effective pairing interaction can be obtained by solving the Bethe-Salpeter equation:
\begin{eqnarray} \label{BSE}
	W_{11'} = \Gamma_{11'} + 
	\frac{1}{\hbar^2\beta}\sum_{2}
	\Gamma_{12}|\bar{\mathcal{G}}_2|^2W_{21'}.
\end{eqnarray}
The obtained $W_{11^\prime}$ then enters into Eq.~(\ref{Eliash03}), in which we express it explicitly as $ W_{11^\prime } \equiv W(\omega_{n}-\omega_{n^\prime}, \bm k-\bm k^\prime)$.  We note that in general $W_{11^\prime}$ depends on $1$ and $1^\prime$ independently.  For EPC, it is a good approximation to regard $W_{11^\prime}$ as a function of $\omega_{n}-\omega_{n^\prime}$ as long as $\hbar|\omega_{n(n^\prime)}|$ is much smaller than the Fermi energy.  Moreover, since liquids are isotropic and only matrix elements with both $\bm k$ and $\bm k^\prime$ residing on the Fermi surface are needed for evaluating Eq.~(\ref{Eliash03}), $W_{11^\prime}$ can be written as a function of $\bm k-\bm k^\prime$.

\subsection{Approximations}\label{G_appro}
To develop an algorithm for estimating $T_c$ based on the the rigorous formalism shown in the last subsection, Liu et al. employ a number of approximations.  In this subsection, we will discuss these approximations, and identify the one to be improved.

The most important approximation which makes an implementation possible is the quasi-static approximation.  It is necessary because solving Eqs.~(\ref{GreenG}) and (\ref{Tmatrix}) exactly  is deemed to be impossible in practice as it requires a time resolution $\sim \hbar/\epsilon_F$ or a corresponding large cutoff frequency $\sim \epsilon_F/\hbar$.  Fortunately, one can exploit the fact that ions move much slower than electrons.  It means that the ionic field only has significant components for frequencies $\nu_m \equiv 2\pi m/\hbar\beta \lesssim \omega_\mathrm{ph}$, where $\omega_\mathrm{ph}$ is a characteristic frequency of phonons. In this case, the equation can be solved by applying the quasi-static approximation. This is to choose a large $n$ such that $\omega_n \gg \omega_\mathrm{ph}$, and solve Eq.~(\ref{GreenG}) as if $\hat{H}(\tau)$ is time-independent:
\begin{eqnarray}\label{Gntau}
	\hat{\mathcal{G}}_n(\tau)
	=\hbar\left[\left(\mathrm{i}\hbar\omega_n+ \epsilon_F\right)\hat{\mathbb{I}}-\hat{H}(\tau)\right]^{-1},
\end{eqnarray}
The ensemble average $\hat{\bar{\mathcal{G}}}_n = \langle\hat{\mathcal{G}}_n(\tau)\rangle_C$ will be independent of $\tau$ and a good approximation for the Fourier transform of $\bar{\mathcal{G}}(\tau-\tau^\prime)$ at the frequency $\omega_n$~\cite{liu_superconducting_2020}. The $\mathcal{T}$-matrix can then be calculated in the time-domain in the same manner, and Fourier transformed back to the frequency domain. See Sec.~\ref{AEGreen} for details. For determining the effective pairing potential, we only need to solve for a specific $n$ with $\omega_\mathrm{ph} \ll \omega_n \ll \epsilon_F/\hbar$.  This is because $W_{11^\prime} \equiv W(\omega_{n}-\omega_{n^\prime}, \bm k-\bm k^\prime)$ is assumed to be a function of  $\omega_n - \omega_n^\prime \equiv \nu_m$.
%Similarly, Eq.~(\ref{LSE}) can solved as if $\hat{\mathcal{V}}(\tau)$ is a static potential:
%\begin{equation}\label{Tntau}
%	\hat{\mathcal{T}}_n(\tau) = \left[\hat{\mathbb{I}}-\frac{1}{\hbar}\hat{\mathcal{V}}(\tau)\hat{\bar{\mathcal{G}}}_n \right]^{-1}\hat{\mathcal{V}}(\tau).
%\end{equation}
%The $\mathcal{T}$-matrix defined by Eq.~(\ref{Tmatrix}) can be approximated by
%\begin{eqnarray}
%	\hat{\mathcal{T}}_{\omega_{n} + \nu_m,\omega_n} \approx \frac{1}{\hbar\beta}\int_{0}^{\hbar\beta}d\tau
%	\hat{\mathcal{T}}_n(\tau)\mathrm{e}^{\mathrm{i}\nu_m\tau}\label{QSA1.1},
%\end{eqnarray}
%with $\nu_m\equiv2\pi m/\hbar\beta$.  For determining the effective pairing potential, we only need to solve for a specific $n$ with $\omega_\mathrm{ph} \ll \omega_n \ll \epsilon_F/\hbar$.  This is because $W_{11^\prime} \equiv W(\omega_{n}-\omega_{n^\prime}, \bm k-\bm k^\prime)$ is assumed to be a function of  $\omega_n - \omega_n^\prime \equiv \nu_m$.

More approximations are involved in PIMD simulations.  It is necessary to simulate a system in a finite size supercell, and discretize the imaginary time to a finite number of beads.  In Ref.~\onlinecite{liu_superconducting_2020}, it is found that the discretization truncates the high-frequency components of the density correlation function of ions and the effective pairing interaction.  It results in an incorrect asymptotic behavior of $\lambda(m)$ at large $m$.  To address the issue, Liu et al. develop an oversampling scheme to interpolate the density correlation function in the temporal domain, and obtain an over-sampled density correlation function. See Ref.~\onlinecite{liu_superconducting_2020} for details.  To get the effective pairing interaction, they make use of the empirical relation:
\begin{equation}
	W(\nu_m,\bm q) = |M(\nu_m,\bm q)|^2 \chi(\nu_m,\bm q),\label{WtoChi}
\end{equation}
where $\chi(\nu_m,\bm q)$ is the density correlation function of ions, and $M(\nu_m,\bm q)$ could be regarded as the EPC matrix element as defined in the Eliashberg theory. It is found numerically that the relation holds well for metallic hydrogen liquids, and $|M(\nu_m,\bm q)|^2$ can be obtained from a linear regression.  By substituting the density correlation function in Eq.~(\ref{WtoChi}) with the over-sampled one, one can obtain an over-sampled effective pairing interaction.  It is found that $\lambda(m)$ determined from the over-sampled effective pairing interaction has a correct asymptotic behavior at large $m$.

Lastly, Liu et al. adopt the LSA for evaluating the ionic field.  This is to approximate the Fourier transform of the ionic potential as
\begin{equation}
	V_\mathrm{ion}(\bm q,\tau) = \frac{e^2}{\varepsilon_\mathrm{et}(q)q^2} \rho_\mathrm{i}(\bm q, \tau),
\end{equation}
where $\rho_\mathrm{i}(\bm q, \tau)=\sum_a \exp[-\mathrm i \bm q \cdot \bm R_a(\tau)]$ is the Fourier transform of the density distribution of ions, and $\varepsilon_\mathrm{et}(q)$ is the static electron-test charge dielectric function~\cite{giuliani_quantum_2005} which has a form similar to that of the random phase approximation but with a local field correction factor determined by Ichimaru et~al.~\cite{ichimaru_analytic_1981}.  It is derived from the jellium model which models ions as a uniform positive charge background, an approximation which obviously breaks down in core regions close to ions.  While it is reasonable to expect that such an approximation could work well for metallic hydrogen in which ions (protons) have small core volumes and electrons are shared by all ions, it becomes problematic when other atoms are involved and bonding between atoms is covalent in nature.

Among the three approximations, the LSA is the one that severely limits the applicability of Liu et al.'s implementation for systems other than metallic hydrogen. %It also disqualifies the implementation as a full first-principles implementation.
This will be the focus of this paper aiming to improve the implementation.  The effect of the finite simulation supercell and the finite number of beads should be assessed to make sure that calculations converge.  This will also be tested.  Finally, it is the quasi-static approximation which makes an implementation feasible for EPC.  The approximation is valid because phonons have an energy scale much smaller than that of electrons, a feature unique to conventional EPC superconductors.

\section{Formalism for the PAW method}\label{PAW2AE}
In this section, we develop formalism for a full DFT-based implementation.  To balance speed and accuracy, we base our implementation on the PAW method \cite{blochl_projector_1994,kresse_ultrasoft_1999}.  As a pseudo-potential method, the PAW method is efficient. Meanwhile, one can obtain AE wavefunctions from PS wavefunctions by applying a transformation. The latter is particularly useful for us since the formalism presented in Sec.~\ref{Theory} is established in an AE basis. To facilitate an implementation based on the PAW method, we need formulae of determining the AE Green's functions and the $\mathcal{T}$-matrices by using quantities defined in the PS basis.

In Sec.~\ref{PAW}, we first briefly review the PAW method and outline its formalism. With the preparation, we derive formulae for determining the AE Green's functions and the $\mathcal{T}$-matrices by using the PAW method in Sec.~\ref{AEGreen}.  In Sec.~\ref{ModifiedOverlap}, we discuss the evaluations of quantities involved in the formulae, including the transformation matrix and the overlap matrix.

\subsection{PAW method}\label{PAW}
In the PAW method, electrons are described by PS wavefunctions. A transformation operator is introduced to map a PS wavefunction $\tilde{\Psi}$ to an AE wave function $\Psi$:
\begin{eqnarray}\label{T1}
	\Psi=\hat{T}\tilde{\Psi}\equiv\tilde{\Psi}+
	\sum_{a}\hat{T}^a|\tilde{\Psi}\rangle,
\end{eqnarray}
where $\hat{T}^a$ is an operator transforming the core state of the ion $a$ from the PS space to the AE space, and is nonzero only within an augmentation sphere surrounding the ion. The radius of the sphere is chosen such that there is no overlap between spheres associated with different ions. Inside a sphere, the wave functions are expanded in the basis of partial waves:
\begin{eqnarray}\label{T2}
	\hat{T}^a|\tilde{\Psi}\rangle=|\Psi^a\rangle-|\tilde{\Psi}^a\rangle
	=\sum_i\left(|\phi_i^a\rangle-|\tilde{\phi}_i^a\rangle\right)\langle\tilde{p}_i^a|\tilde{\Psi}\rangle,
	\quad
\end{eqnarray}
where $i$ indexes the partial wave states with angular momentum quantum numbers $(l_i m_i)$ and at a given reference energy $\epsilon_{i}$. The AE partial wavefunctions $\langle\bm{r}|\phi_i^a\rangle\equiv\phi_{l_i \epsilon_i}^a(r)Y_{l_i m_i}(\hat{\bm{r}})$ at the radial coordinate $r$ and along the direction $\hat{\bm{r}}$ are obtained by solving a spherical AE Schrodinger equation at the reference energy in the presence of an ion at the center. The corresponding PS partial wavefunctions $\langle\bm{r}|\tilde{\phi}_i^a\rangle$ are chosen such that they are smooth and coincide with the AE partial wavefunctions outside the sphere. $\langle \tilde{p}_i^a|$ are a set of projectors which are biorthogonal to the PS partial wave states:
\begin{eqnarray}
	\langle\tilde{p}_j^a|\tilde{\phi}_i^a\rangle=\delta_{i,j}.
\end{eqnarray}
With a complete set of the basis, the closure relation in each sphere can be written as:
\begin{eqnarray}\label{closure}
	\sum_i|\tilde{\phi}_i^a\rangle\langle\tilde{p}_i^a|=\hat{\mathbb{I}}^a.
\end{eqnarray}

In the PS basis, the Kohn-Sham (KS) equation is transformed to 
\begin{subequations}\label{eigen}
	\begin{eqnarray}
		\hat{\tilde{H}}\tilde{\Psi}_n&=&\epsilon_n\hat{S}\tilde{\Psi}_n,\label{eigeneq}\\
		\hat{\tilde{H}}&=&\hat{T}^\dagger\hat{H}\hat{T}\label{PSH},\\
		\hat{S}&=&\hat{T}^\dagger\hat{T}\label{Overlap_ori},
	\end{eqnarray}
\end{subequations}
where $\hat{H}$ and $\hat{\tilde{H}}$ denote the Hamiltonians in the AE and PS basis, respectively.  This is a generalized eigenvalue problem, and the orthogonality condition between PS eigen-wavefunctions becomes
\begin{eqnarray}\label{orthonormal}
	\langle\Psi_m|\Psi_n\rangle
	=\langle\tilde{\Psi}_m|\hat{S}|\tilde{\Psi}_n\rangle
	=\delta_{m,n}.
\end{eqnarray}
We note that in DFT calculations based on the PAW method, the PS Hamiltonian is directly derived from a total energy functional expressed in terms of the PS wavefunctions.  It is not necessary to construct the AE Hamiltonian first and then transform it to the PS Hamiltonian by using Eq.~(\ref{PSH}).
% The Hamiltonian can finally be expressed in a simple form:
% \begin{eqnarray}
% 	\hat{\tilde{H}}=-\frac{1}{2}\nabla^2+\tilde{v}_{eff}
% 	+\sum_{ij,a}|\tilde{p}_i^a\rangle (\tilde{v}_{nl})_{ij}^a \langle\tilde{p}_j^a|,
% \end{eqnarray}
% where $\tilde{\hat{v}}_{nl}$ is the non-local potential introduced by PAW method to recover the correct properties of AE wave functions.

\subsection{AE Green's function and $\mathcal{T}$-matrix}\label{AEGreen}
With the preparation, we now proceed to derive formulae for determining the AE Green's function and the $\mathcal{T}$-matrix by using the PAW method.

We seek for an expression of the Green's function appropriate for the PAW method. Under the quasi-static approximation, the AE Green's function is determined by Eq.~(\ref{Gntau}). However, it depends on the AE Hamiltonian $\hat{H}(\tau)$. We can transform it to a form that only depends on quantities in the PS basis.  To see that, we start from the spectral representation of a general Green's function $\hat{\mathcal{G}}(\epsilon)= \hbar(\epsilon\hat{\mathbb{I}}-\hat{H})^{-1}$:
\begin{eqnarray}
	\hat{\mathcal{G}}(\epsilon)
	&=&\hbar\sum_{n}\frac{|\Psi_{n}\rangle\langle\Psi_{n}|}{\epsilon-\epsilon_n}\nonumber\\
	&=&\hat{T}
	\left[\hbar\sum_{n}\frac{|\tilde{\Psi}_{n}\rangle\langle\tilde{\Psi}_{n}|}{\epsilon-\epsilon_n}\right]
	\hat{T}^\dagger, \label{GTT}
\end{eqnarray}
where $|\Psi_n\rangle$ and $|\tilde{\Psi}_n\rangle$ are the AE and PS eigen-wavefunctions, respectively, and $\epsilon_n$ is the corresponding eigen-energy.

We can convert the expression to a more convenient form. By using the orthonormality condition Eq.~(\ref{orthonormal}), it is easy to verify the closure relation
\begin{eqnarray}
	\sum_n\hat{S}|\tilde{\Psi}_{n}\rangle\langle\tilde{\Psi}_{n}|=\hat{\mathbb{I}}.
\end{eqnarray}
We then have
\begin{eqnarray}
	\left(\epsilon\hat{S}-\hat{\tilde{H}}\right)^{-1}
	=&&\sum_{n}\left(\epsilon\hat{S}-\hat{\tilde{H}}\right)^{-1}
	\hat{S}|\tilde{\Psi}_{n}\rangle\langle\tilde{\Psi}_{n}|\nonumber\\
	=&&\sum_{n}\frac{|\tilde{\Psi}_{n}\rangle\langle\tilde{\Psi}_{n}|}{\epsilon-\epsilon_n},
\end{eqnarray}
where we make use of the eigen-equation Eq.~(\ref{eigeneq}). By substituting the equality into Eq.~(\ref{GTT}) and setting $\epsilon = \mathrm{i}\hbar\omega_n +\epsilon_F$, we obtain the formula of determining the AE Green's function in the PAW method:
\begin{equation}\label{QSA1.4}
	\hat{\mathcal{G}}_n(\tau)=\hbar\hat{T}(\tau)\left[(\mathrm{i}\hbar\omega_n+\epsilon_F)\hat{S}(\tau)-\hat{\tilde{H}}(\tau)\right]^{-1}\hat{T}^\dagger(\tau).\quad
\end{equation}
We note that $\hat{\tilde{H}}$, $\hat{S}$ and $\hat{T}$ all depend on the configuration of ions, therefore are time-dependent.

With the Green's function, the scattering $\mathcal{T}$-matrix can be determined. To avoid solving the Lippmann-Schwinger equation in the AE space, we apply an identity 
\begin{equation}
	\hat{\mathcal{T}}_n(\tau)=\hbar\hat{\bar{\mathcal{G}}}_n^{-1}\hat{\mathcal{G}}(\tau)\hat{\bar{\mathcal{G}}}_n^{-1}-\hbar\hat{\bar{\mathcal{G}}}_n^{-1}\label{Tmatrix}
\end{equation}
to calculate the $\mathcal{T}$-matirx in time domain. It is easy to prove that the $\mathcal{T}$-matirx calculated in this way satisfies Eq.~(\ref{LSE}). We note that in a liquid, because of the spatial translation symmetry, the average Green's function is diagonal in the plane-wave basis. As a result, the matrix inverse and multiplication can be carried out in a truncated Hilbert space spanned by the basis without introducing errors.

The matrix elements of $\mathcal{T}$-matrix in frequency domain can then be determined by applying the Fourier transform:
\begin{eqnarray}
	\hat{\mathcal{T}}_{\omega_{n} + \nu_m,\omega_n} \approx \frac{1}{\hbar\beta}\int_{0}^{\hbar\beta}d\tau
	\hat{\mathcal{T}}_n(\tau)\mathrm{e}^{\mathrm{i}\nu_m\tau}\label{QSA1.1}.
\end{eqnarray}
The matrix elements are all that are needed for evaluating the pair scattering amplitude and the effective pairing interaction (see Sec.~\ref{G_theory}).  We note that for the purpose only the matrix elements with wave-vectors close to the Fermi surface are needed.  Therefore, a large energy cutoff, which is required for recovering the AE wavefunctions from the PS wavefunctions, is not needed for our calculation.

\subsection{Transformation and overlap matrices}\label{ModifiedOverlap}
To apply the formulae derived in the last subsection, it is necessary to determine the transformation operator $\hat{T}$ and the overlap matrix $\hat{S}$.  

The transformation operator transforms a PS state to an AE state. From Eq.~(\ref{T1}) and Eq.~(\ref{T2}), the operator is
\begin{equation}\label{Top}
	\hat{T}=\hat{\mathbb{I}}+{\sum_{j,a}}'\left(|\phi_j^a\rangle-|\tilde{\phi}_j^a\rangle\right)\langle\tilde{p}_j^a|,
\end{equation}
where the summation is over ions and partial waves, with the prime indicating explicitly that the summation is always truncated to a small number of partial waves in real calculations.  For liquids, the basis states in both the AE space and the PS space are the plane waves.  A matrix element in the plane wave basis is
\begin{multline}
	\langle\psi_{\bm{k}}|\hat{T}|\tilde{\psi}_{\bm{k}'}\rangle =
	\delta_{\bm k \bm k^\prime}\\
	+{\sum_{i,a}}'\left(\langle\psi_{\bm{k}}|\phi_i^a\rangle
	-\langle\psi_{\bm{k}}|\tilde{\phi}_i^a\rangle\right)\langle\tilde{p}_i^a|\tilde{\psi}_{\bm{k}'}\rangle , \label{Trans}
\end{multline}
where $|\psi_{\bm{k}}\rangle$ and $|\tilde{\psi}_{\bm{k}^\prime}\rangle$ are the plane waves in the AE and PS space, respectively. The overlap matrix element between a plane wave and a partial wave can be evaluated by
\begin{multline}
	\langle\psi_{\bm{k}}|\phi_i^a\rangle
	=\frac{4\pi}{\sqrt{V}}\mathrm{e}^{-\mathrm{i}\bm{k}\cdot\bm{R}_a} Y_{l_i m_i}(\hat{\bm{k}})\\
	\times\int_{0}^{r_c}r^2\mathrm{d}r(-\mathrm{i})^{l_i} j_{l_i}(kr)\phi_{l_i,m_i}^a(r), \label{kphi}
\end{multline}
and similarly for other matrix elements, where $Y_{lm}$ denotes the spherical harmonics,  $j_l(kr)$ is the spherical Bessel function, $r_c$ denotes the cutoff radius of the augmentation sphere in defining $\hat{T}^a$, and $\bm{R}_a$ is the position of the ion $a$.

The overlap matrix $\hat{S}=\hat{T}^\dagger\hat{T}$ can also be determined by using the partial waves.  From Eq.~(\ref{Top}), we have
\begin{multline}\label{Overlap}
	\hat{S}=-\hat{\mathbb{I}}
	+\hat{T} + \hat{T}^\dagger \\
	+{\sum_{ij,a}}'\Ket{\tilde{p}_i^a}
	\Braket{\phi_i^a -\tilde\phi_i^a|\phi_j^a - \tilde \phi_j^a}
	\Bra{\tilde{p}_j^a},
\end{multline}
where we assume that augmentation spheres associating with different ions do not overlap.  We can rewrite it as
\begin{equation}\label{Overlap2}
	\hat{S} = \hat{S}^\prime+\left(\hat{T}-\hat{T}^\prime+\mathrm{h.c.}\right),
\end{equation}
where
\begin{eqnarray}
	\hat{S}^\prime &=& \hat{\mathbb{I}}+{\sum_{ij,a}}'
	|\tilde{p}_i^a\rangle
	\left(\langle\phi_i^a|\phi_j^a\rangle-\langle\tilde{\phi}_i^a|\tilde{\phi}_j^a\rangle\right)
	\langle\tilde{p}_j^a|, \label{Overlap1} \\
	\hat{T}^\prime &=& \hat{\mathbb{I}}+{\sum_{ij,a}}'|\tilde{p}_j^a\rangle\left(\langle\tilde{\phi}_j^a|\phi_i^a\rangle
	-\langle\tilde{\phi}_j^a|\tilde{\phi}_i^a\rangle\right)\langle\tilde{p}_i^a|.\label{Trans1}
\end{eqnarray}
We note that $\hat{S}^\prime$ is the form commonly adopted in literatures.  Actually, by using the closure relation Eq.~(\ref{closure}), it is easy to see that $\hat{T}^\prime$ is equal to $\hat{T}$, and therefore $\hat{S}=\hat{S}^\prime$ if the summation is over a complete set of partial waves.  However, since real calculations always use just a small set of partial waves instead of a complete set, the correction due to $\hat{T} - \hat{T}^\prime$ could be non-negligible.  For the determination of the AE Green's function and the $\mathcal{T}$-matrix, it is important to maintain the consistency between $\hat{T}$ and $\hat{S}$ since they both appear in Eq.~(\ref{QSA1.4}).  For the reason, we use Eq.~(\ref{Overlap2}) instead of Eq.~(\ref{Overlap1}) for calculating the overlap matrix.

% Overlap between augmentation spheres also introduces error. Especially for liquids, overlap between some of the spheres is almost inevitable. However, the error involves integrations between different spherical grids and thus has a very complicated form. G. Kresse suggests that with their method of constructing partial waves, and the overlap will lead only to small error. As for liquid Hydrogen, we perform calculations with PAW datasets of different cutoff radii to check the influence of overlap. The results only show little difference.

\section{Implementation and results}

\subsection{Implementation}

\begin{figure}
	\includegraphics[width=8.6cm]{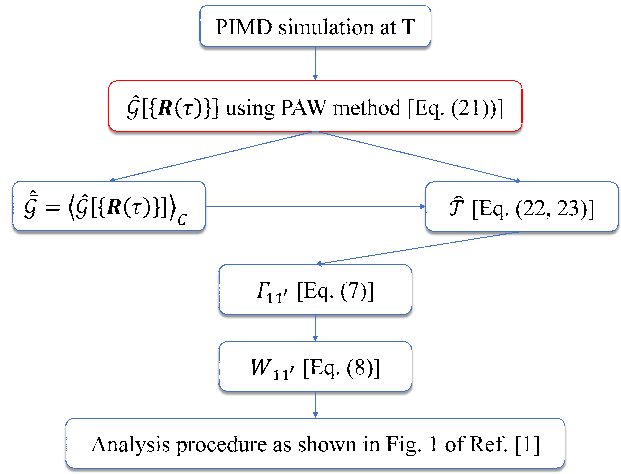}
	\caption{\label{fig:flow} Flowchart for calculating Green's function and $\mathcal{T}$-matrix. It substitutes the corresponding part of the workflow shown in Fig.~1 of Ref.~\cite{liu_superconducting_2020}.}
\end{figure}

We employ the GPU version of VASP~\cite{hacene_accelerating_2012,hutchinson_vasp_2012} for calculating electron structures. The radius of the augmentation spheres is set to $0.52$\,\AA, which may result in overlaps of the spheres in some ionic configurations but saves computation time. We  test a smaller radius of $0.43$\,\AA\ and find that differences in calculation results are negligible. The Perdew-Burke-Ernzerhof (PBE) exchange-correlation functional~\cite{perdew_generalized_1996} is used in all calculations. A Monkhorst-Pack k-point mesh of spacing not larger than $2\pi\times0.05$\,\AA$^{-1}$ is used to sample the Brillouin zone, and an energy cutoff of $600$\,eV is employed to expand the wavefunctions, yielding a convergence in energy better than $\sim 2\, \mathrm{meV/atom}$.

To determine the Green's function and the scattering $\mathcal{T}$-matrix by using Eqs.~(\ref{QSA1.4}, \ref{Tmatrix}), we need to obtain the pseudo Hamiltonian $\hat{\tilde{H}}$, the transformation matrix $\hat{T}$ and the overlap matrix $\hat{S}$ from the electron structure calculations. Since VASP does not calculate or output these quantities directly, we modify it to output intermediate quantities including the local part of the effective KS potential $\tilde{v}_{\mathrm{loc}}$,  the coefficients of the non-local part of the effective KS potential $(\tilde{v}^\prime)_{ij}^a$ as well as the projector matrix elements $\langle\tilde\psi_{\bm{k}} |\tilde{p}_i^a \rangle$.  With them, the matrix elements $\langle\tilde\psi_{\bm k}| \hat{\tilde H} |\tilde\psi_{\bm k^\prime}\rangle$ of the pseudo Hamiltonian can be determined by using the definition
\begin{eqnarray}
	\hat{\tilde{H}}=\hat{K}+\tilde{v}_{\mathrm{loc}}
	+\sum_{ij,a}|\tilde{p}_i^a\rangle (\tilde{v}^\prime)_{ij}^a \langle\tilde{p}_j^a|.
\end{eqnarray}
The matrix elements of the transformation operators $\hat{T}$, $\hat{T}^\prime$ and the overlap matrix $\hat{S}'$ can also be determined by applying Eq.~(\ref{Top}), Eq.~(\ref{Trans1}), and Eq.~(\ref{Overlap1}), respectively.  It is then straightforward to determine $\hat S$ by applying Eq.~(\ref{Overlap2}).

We can then proceed to carry out the analysis outlined in Sec.~\ref{Theory} by following the procedure detailed in Ref.~\onlinecite{liu_superconducting_2020}.  Figure~\ref{fig:flow} shows the flowchart of our implementation.

\subsection{Results}\label{Result}

\begin{figure}[t]
	\includegraphics[width=8.6cm]{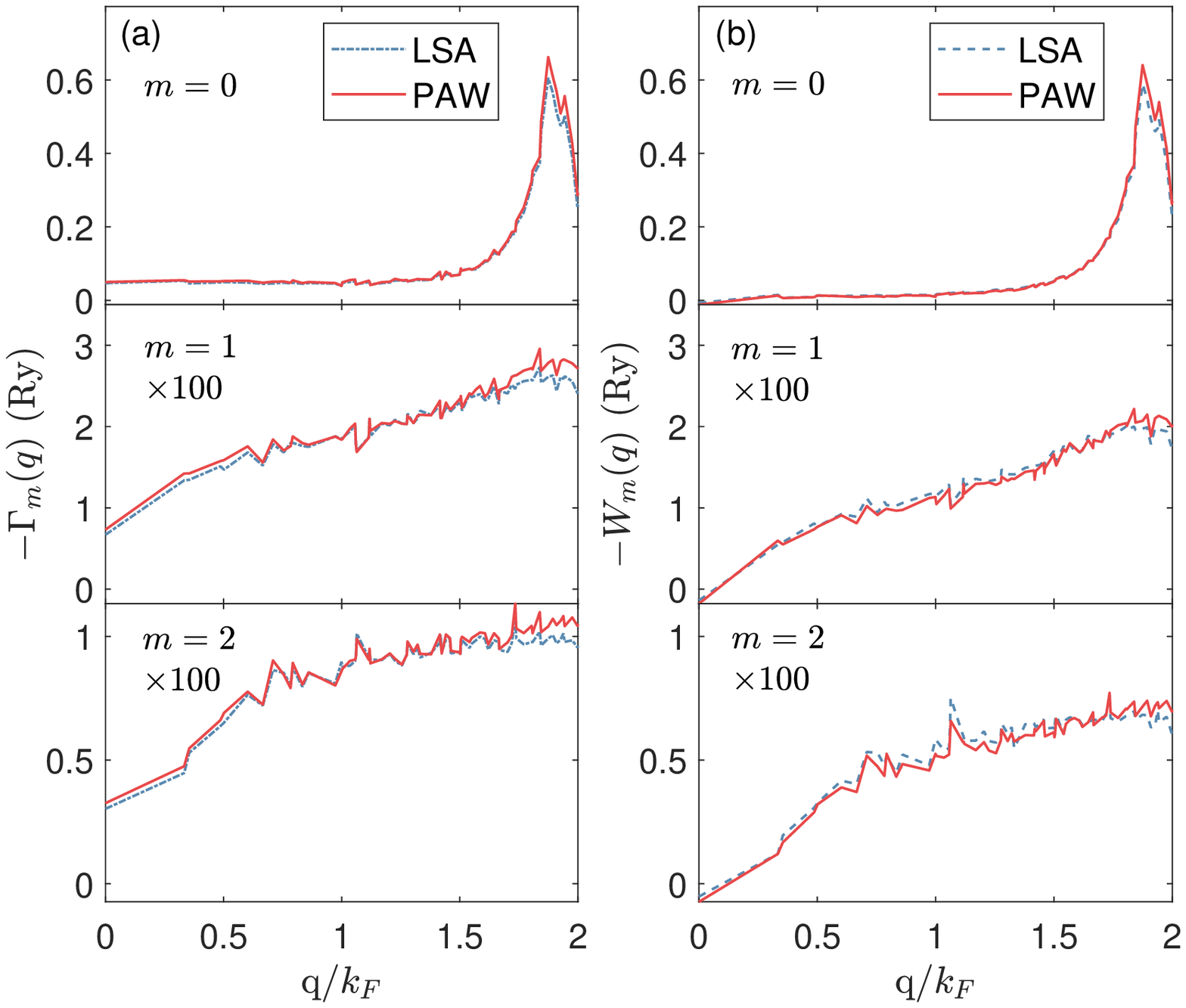}
	\includegraphics[width=8.6cm]{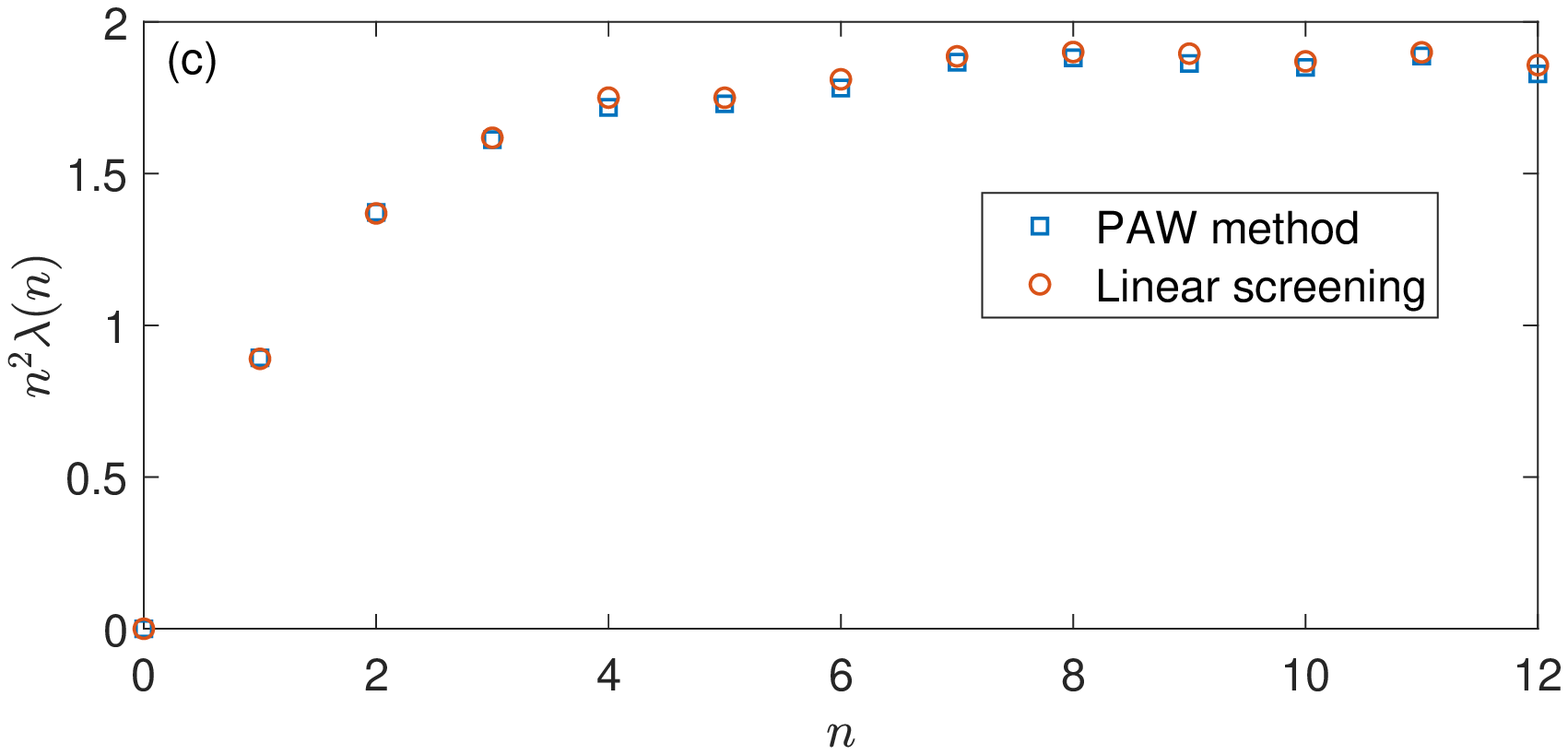}
	\caption{\label{fig:Gamma} Comparison between the full DFT-based implementation (PAW method) and the LSA in (a) pair scattering amplitude $\Gamma_m(q)$, (b) effective pair interaction $W_m(q)$ and (c) interaction parameters $\lambda(n)$.  The results are  for $T=450$\,K and $P=500$\,GPa. $\Gamma_m(q)$ and $W_m(q)$ are obtained by recasting the matrix elements of $\Gamma$ and $W$ as a function of $q = |\bm k_1 - \bm k_2|$ for $\bm k_1$, $\bm k_2$ in $0.8k_F\le|\bm{k}_1|,|\bm{k}_2|\le1.14k_F$.}
\end{figure}

To test our first-principles implementation based on PAW method, we apply it to metallic hydrogen liquids. The ion configurations from the constant-volume PIMD simulations of 200 hydrogen atoms in Ref.~\cite{liu_superconducting_2020} are re-used in the current study. In this subsection, we summarize our calculation results.

Estimated $T_c$'s as well as relevant parameters of metallic hydrogen liquids under pressures from $0.5$ to $1.5$\,TPa are summarized in Table~\ref{tab:result}. For comparison, the LSA results are also shown.  We find that the results coincide well with the LSA results. It confirms Liu et~al.'s prediction that the superconducting transition temperature of metallic hydrogen is higher than its predicted melting temperature~\cite{liu_superconducting_2020}.  We note that our prediction of $T_c$ is only valid for the liquid phase. $T_c$ for the solid phase of metallic hydrogen had been thoroughly studied in Ref.~\cite{mcmahon_high-temperature_2011}.
%  The isotope effect is checked by performing a calculation for liquid metallic deuterium under $1000GPa$. The predicted $T_c=283K$ is close to that suggested by isotope effect, where $418K/\sqrt{2}\approx295K$.

\begin{figure}[t]
	\includegraphics[width=8.6cm]{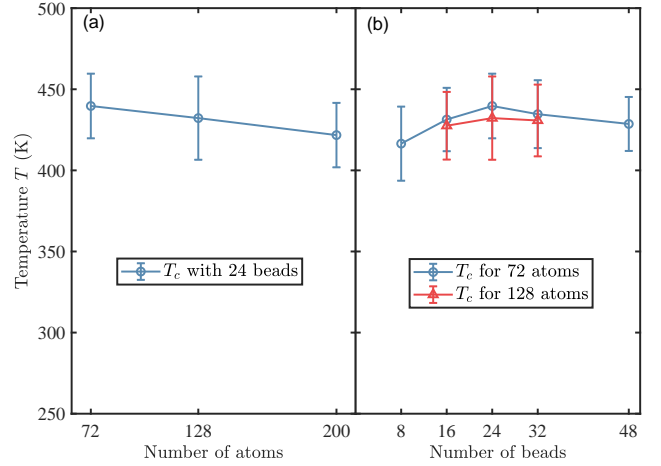}% Here is how to import EPS art
	\caption{\label{fig:convergence} Dependence of $T_c$ on (a) different numbers of atoms, and (b) different numbers of beads for $P=700$\,GPa. The error bars indicate the numerical uncertainties of the estimations of $T_c$'s, i.e., the values shown in the parentheses in Table~\ref{tab:result}.}
\end{figure}

In Fig.~\ref{fig:Gamma}, we show the comparison of the pair scattering amplitude $\Gamma_m(q)$ and the effective pairing interaction $W_m(q,\nu)$ between the two implementations for $P=700$\,GPa and $T=450$\,K.  We find that the results yielded by the two implementations are very close. We also show the interaction parameters $\lambda(n)$ in Fig.~\ref{fig:Gamma}(c). To determine $\lambda(n)$, we recast the matrix element $W_{11^\prime}$ as a function of $q=|\bm k - \bm k^\prime|$ for $0.8k_F\le|\bm{k}|,|\bm{k}^\prime|\le1.14k_F$, and evaluate Eq.~(\ref{Eliash03}) by integrating over $q$~\cite{jaffe_superconductivity_1981}.  It is evident that the results also coincide well.  It is not surprising that the LSA can work well in metallic hydrogen systems because hydrogen has no inner-core electron and the effective KS potential is dominated by its local part which can be well approximated by the LSA with an appropriate dielectric function.  For more general systems like hydrides, the LSA is obviously inadequate, and our implementation based on the standard first-principles approach will be useful. 

% We summarize our results for metallic hydrogen liquid in this section. As suggested by Eq.~(\ref{WtoChi}), we recast the effective pairing interaction $W(\bm{k}_1,\omega_{N_s};\bm{k}_2,\omega_{N_s}+\nu_m)$ as a function of $q=|\bm{k}_1-\bm{k}_2|$ and $\nu_m$, for $\bm{k}_1$ and $\bm{k}_2$ near the Fermi surface. It is then fitted to the effective EPC matrix model. In Fig.~\ref{fig:main}, we plot $W_m(q)$ for hydrogen liquid under pressure of $700GPa$ at $350K$. The fitted EPC matrix $M_m(q)$ in shown in the inset. It can be seen that the model fits the first-principle results well.
% \begin{figure}
% 	\includegraphics[width=8.6cm]{main_700_350}
% 	\caption{\label{fig:main} Effective pairing interaction $W_m(q)$ for $m=0,1,2$ components, fitted to the effective EPC matrix model $W_m(q)=|M_m(q)|^2\chi_i(q,\nu_m)$ from matrix elements around the Fermi surface $0.8k_F\le|\bm{k}_1|,|\bm{k}_2|\le1.14k_F$. The calculation is performed with $N_s = 16$ for a 200-atom supercell under the pressure of $P = 700 GPa$ at $T = 350K$. The scatter points and error bars show the averages and standard deviations of $W_m(q)$ with the same $q$ but different $\bm{k}_1$’s and $\bm{k}_2$’s. Inset: The effective squared EPC matrix elements $\beta\rho_0|M_m(q)|^2$, where $\rho_0$ is the ion density.}
% \end{figure}

We test the convergence of our calculation by varying the size of the simulation supercell and the number of beads discretizing the imaginary time.  In Ref.~\onlinecite{liu_superconducting_2020} and the calculation shown above, we use a supercell with 200 hydrogen atoms and discretize the imaginary time to 24 beads. In Fig.~\ref{fig:convergence}(a), we show the dependence of estimated $T_c$ on the number of atoms. We find that $T_c$ slightly decreases with the increasing simulation size,  but the difference are within the error bars.  In Fig.~\ref{fig:convergence}(b), we show the dependence of estimated $T_c$ on the number of beads.  We find that only a relatively small number of beads ($>16$) is needed to get a converged result. The oversampling procedure discussed in Sec.~\ref{G_appro} is effective in eliminating the discretization errors.

\begin{table*}%The best place to locate the table environment is directly after its first reference in text
	\begin{ruledtabular}
		\renewcommand{\arraystretch}{1.4}
		\begin{tabular}{lcccccccc}
			& \multicolumn{2}{c}{$r_s$}                 & $1.226$                          & $1.197$     & $1.17$       & $1.149$     & $1.113$     & $1.049$                   \\
			& \multicolumn{2}{c}{P}                 & $0.5$                            & $0.6$       & $0.7$       & $0.8$       & $1.0$       & $1.5$                   \\
			\colrule
			\multirow{7}{*}{\rotatebox{90}{\shortstack{PAW method}}}
			& \multirow{3}{*}{\rotatebox{90}{350K}} & \textrm{$\lambda$}             & $10.2(15)$  & $9.0(10)$   & $8.5(10)$   & $7.5(9)$    & $6.3(8)$    & $5.2(5)$    \\
			&                                       & \textrm{$\overline{\omega}_2$} & $102.1(10)$ & $110.9(10)$ & $112.2(15)$ & $124.6(7)$  & $134.6(3)$  & $164.1(30)$ \\
			&                                       & \textrm{$\rho_m$}              & $0.37(12)$  & $0.36(11)$  & $0.32(10)$  & $0.31(9)$   & $0.32(11)$  & $0.26(9)$   \\
			\cline{2-9}
			& \multirow{3}{*}{\rotatebox{90}{450K}} & \textrm{$\lambda$}             & $8.2(16)$   & $7.7(10)$   & $7.1(8)$    & $6.3(10)$   & $5.5(6)$    & $4.6(4)$    \\
			&                                       & \textrm{$\overline{\omega}_2$} & $115.0(1)$  & $114.9(13)$ & $132.0(7)$  & $141.9(4)$  & $147.3(11)$ & $173.1(12)$ \\
			&                                       & \textrm{$\rho_m$}              & $-0.10(12)$ & $-0.13(9)$  & $-0.13(8)$  & $-0.10(10)$ & $-0.14(7)$  & $-0.14(6)$  \\
			\cline{2-9}
			& \multicolumn{2}{c}{$T_c$}             & $428(25)$                        & $423(19)$   & $421(19)$   & $425(24)$   & $420(19)$   & $415(17)$               \\
			\hline\hline
			\multirow{7}{*}{\rotatebox{90}{\shortstack{Linear screening}}}
			& \multirow{3}{*}{\rotatebox{90}{350K}} & \textrm{$\lambda$}             & $9.5(14)$   & $8.6(11)$   & $8.1(10)$   & $7.2(10)$   & $6.0(8)$    & $5.0(5)$    \\
			&                                       & \textrm{$\overline{\omega}_2$} & $106.7(10)$ & $113.8(1)$  & $115.8(7)$  & $127.5(16)$ & $138.0(1)$  & $167.5(30)$ \\
			&                                       & \textrm{$\rho_m$}              & $0.38(12)$  & $0.36(10)$  & $0.32(10)$  & $0.30(8)$   & $0.29(11)$  & $0.24(9)$   \\
			\cline{2-9}
			& \multirow{3}{*}{\rotatebox{90}{450K}} & \textrm{$\lambda$}             & $7.6(13)$   & $7.4(10)$   & $6.6(8)$    & $6.0(10)$   & $5.2(6)$    & $4.4(4)$    \\
			&                                       & \textrm{$\overline{\omega}_2$} & $119.9(11)$ & $117.8(4)$  & $135.8(3)$  & $144.6(19)$ & $157.1(6)$  & $176.3(12)$ \\
			&                                       & \textrm{$\rho_m$}              & $-0.11(11)$ & $-0.13(9)$  & $-0.14(8)$  & $-0.11(9)$  & $-0.14(7)$  & $-0.15(6)$  \\
			\cline{2-9}
			& \multicolumn{2}{c}{$T_c$}             & $428(24)$                        & $423(19)$   & $420(19)$   & $424(22)$   & $418(19)$   & $411(17)$               \\
		\end{tabular}
	\end{ruledtabular}
	\caption{\label{tab:result}%
		Mass enhancement factor $\lambda\equiv\lambda(0)$,  average phonon frequency $\bar{\omega}_2$ (in meV) and the maximal eigenvalue $\rho_m$ of the linearized Eliashberg equations for different pressures $P$ (in TPa) and temperatures $T$ (in K), determined by the standard first-principles approach (PAW method) and the LSA~\footnote{We recalculate the LSA results. They are slightly different from those presented in Ref.~\cite{liu_superconducting_2020} but within the error bars.  The numerical differences are due to different settings in analyzing the PIMD data.}. The transition temperature $T_c$, i.e., the temperature at which $\rho_m$ becomes zero, is estimated based on a linear interpolation of $\rho_m$ between the two temperatures. Numerical uncertainties of the quantities are indicated in parentheses.}
\end{table*}

\section{Summary}\label{Summary}
In summary, we present a full DFT-based implementation of the stochastic path-integral approach for estimating superconducting transition temperatures. We derive the formulae for determining the Green's function and the scattering $\mathcal{T}$-matrix in the pseudo-basis of the PAW method. Properly calculating the $S$ matrix elements to eliminate the error due to the incompleteness of the basis is also discussed.  We test our implementation in metallic hydrogen liquid systems and get results close to the those obtained in Ref.~\onlinecite{liu_superconducting_2020}.  The calculation is in agreement with Liu et~al's prediction, and suggests that metallic hydrogen could be a superconducting liquid as its superconducting $T_c$ is higher than the predicted melting temperature of metallic hydrogen. With the full DFT-based implementation, it becomes possible to apply the stochastic path-integral approach to a wider class of systems such as hydrides. %The PAW-method-based implementation opens the way for applying the stochastic path-integral approach to a wider class of systems such as hydrides.

This work is supported by the National Basic Research Program of China (973 Program) Grants No. 2018YFA0305603 and No. 2015CB921101 and the National Science Foundation of China Grant No. 11325416.

\bibliographystyle{apsrev4-2}
\bibliography{Reference}

\end{document}